\documentclass[letterpaper]{article} 
\usepackage{aaai25}  
\usepackage{times}  
\usepackage{helvet}  
\usepackage{courier}  
\usepackage[hyphens]{url}  
\usepackage{graphicx} 
\usepackage{natbib}  
\usepackage{caption} 
\usepackage{algorithm}
\usepackage{algorithmic}

\usepackage{newfloat}
\usepackage{listings}
\DeclareCaptionStyle{ruled}{labelfont=normalfont,labelsep=colon,strut=off} 
\floatstyle{ruled}
\newfloat{listing}{tb}{lst}{}
\floatname{listing}{Listing}
\usepackage{subcaption}
\usepackage{amsmath}
\usepackage{amssymb}
\usepackage{dsfont}
\usepackage{booktabs}
\usepackage{multirow}
\usepackage{pifont}

\newcommand{\name}{\textsc{CMSteg}}

\title{Provably Secure Robust Image Steganography via Cross-Modal Error Correction}
\author{
    Yuang Qi,
    Kejiang Chen\thanks{Coresponding author.},
    Na Zhao,
    Zijin Yang,
    Weiming Zhang
}
\affiliations{
    Anhui Province Key Laboratory of Digital Security,\\
    University of Science and Technology of China, Hefei, Anhui, China\\
    \{qiyuang@mail., chenkj@, znzhaona@mail., bsmhmmlf@mail., zhangwm@\}ustc.edu.cn
}

\usepackage{bibentry}

\begin{document}

\maketitle

\begin{abstract}
The rapid development of image generation models has facilitated the widespread dissemination of generated images on social networks, creating favorable conditions for provably secure image steganography. 
However, existing methods face issues such as low quality of generated images and lack of semantic control in the generation process.
To leverage provably secure steganography with more effective and high-performance image generation models, and to ensure that stego images can accurately extract secret messages even after being uploaded to social networks and subjected to lossy processing such as JPEG compression, we propose a high-quality, provably secure, and robust image steganography method based on state-of-the-art autoregressive (AR) image generation models using Vector-Quantized (VQ) tokenizers. 
Additionally, we employ a cross-modal error-correction framework that generates stego text from stego images to aid in restoring lossy images, ultimately enabling the extraction of secret messages embedded within the images. Extensive experiments have demonstrated that the proposed method provides advantages in stego quality, embedding capacity, and robustness, while ensuring provable undetectability.

\end{abstract}

%

\section{Introduction}


Steganography~\cite{cachin2005digital} is a science and art of covert communication that hides secret messages in covers, which needs to avoid arousing suspicion from steganalysis. 
In terms of security, steganography is divided into empirically secure steganography (ESS) and provably secure steganography (PSS). While empirically secure steganography has been developed for many years~\cite{sedighi2015content,wang2019non,wang2020bbc}, there is relatively little research on PSS.
Actually, PSS also has a long history. Cachin~\shortcite{cachin1998information} and Hopper et al.~\shortcite{hopper2002provably} have proposed the definitions of information-theoretic security and computational security for steganography, respectively.

For a long time, PSS has been lacking due to the lack of precise samplers and the inability to obtain a definite cover distribution.
It was not until the emergence of generative artificial intelligence that efficient, efficient PSS became possible~\cite{chen2018provably}, where generative image steganography was in the vanguard.
The image generation model gives an explicit distribution of pixels~\cite{van2016conditional,tulsiani2021pixeltransformer}, or a sampler corresponding to the distribution~\cite{song2019generative,goodfellow2020generative}, which meets the requirements of PSS.


\begin{figure}[t]
    \centering
    \includegraphics[width=\columnwidth]{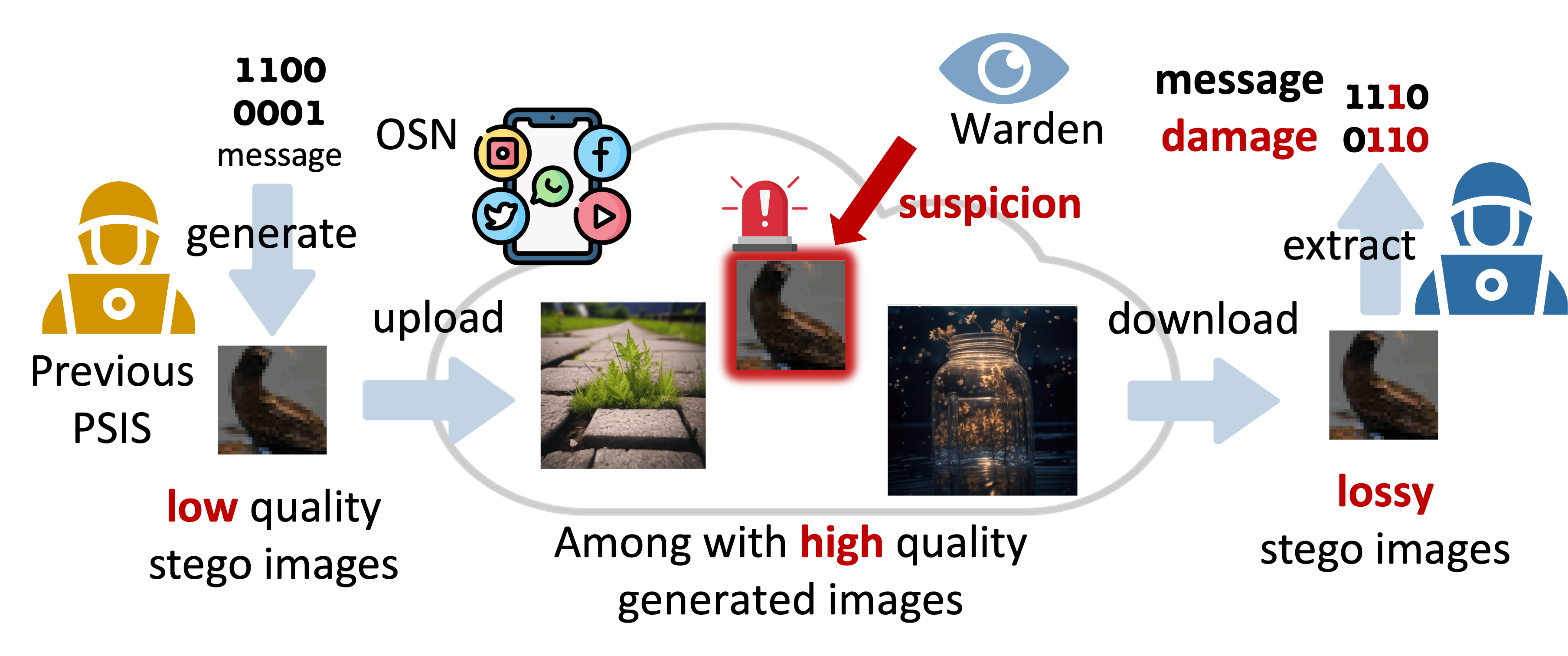}
    \caption{Provably Secure Image Steganography (PSIS) faces challenges in actual transmission within online social networks (OSNs). 
}
    \label{fig:low-resolution}
\end{figure}

Yang et al.~\shortcite{yang2018provably} proposed the first provably secure steganography method based on image generative model, utilizing PixelCNN~\cite{van2016pixel} for message embedding.
Ding et al.~\shortcite{ding2023discop} proposed a provably secure steganography contruction based on distribution copies and deployed it on ImageGPT~\cite{chen2020generative}. 
These two methods can only perform steganography at the pixel level, resulting in stego images with low resolution and poor quality.
In the context where high-resolution image generation models have become increasingly widespread~\cite{zhang2022styleswin,du2024demofusion}, transmitting such low-resolution generated images is no longer an entirely innocent act; this does not align with the covert pursuit of steganographic behavior, as depicted in Figure~\ref{fig:low-resolution}.

Additionally, in practical applications, digital images are widely disseminated through social networks, and steganographic images are no exception. Therefore, the ability to withstand lossy processing by social networks is also an important criterion for evaluating image steganography.
Unfortunately, for the aforementioned provably secure image steganography method, lossy processing can cause the receiver to lose synchronization, leading to heavy message damage.

To resist lossy processing, Yang et al.~\shortcite{yang2023provably} proposed PARIS, a provably secure robust image steganography method using inverse sampling based on generative adversarial networks (GANs), where the message is encoded into a latent vector, and then generating the stego image. GAN inversion~\cite{xia2022gan} is utilized to reconstruct the latent vector and then extract the secret message.
As the GAN network structure deepens, the inversion accuracy decreases rapidly and the message is difficult to extract. Therefore, the inversion-based method can only be limited to low-quality small GAN. 

Su et al. proposed StegaStyleGAN, achieving provably security and higher resolution~\cite{su2024stegastylegan}. They used a message mapping method similar to that of in PARIS to map the message into random noise of StyleGAN~\cite{karras2019style}, and trained a CNN for message extraction. Although StegStyleGAN has the capability to generate stegos with resolutions of $256\times256$, it is specifically designed for StyleGAN, and the quality of the image is limited by the upper bounds of GAN's generative capabilities. 

Large language models (LLMs) offer remarkable performance in solving language tasks~\cite{vaswani2017attention,radford2019language,achiam2023gpt} and showing potential towards achieving general artificial intelligence~\cite{ge2024openagi,almeida2024exploring}, which inspired researchers to explore the possibility of developing autoregressive (AR) models in the field of image generation. 
AR image generation models, represented by Vector-Quantized-VAE (VQ-VAE)~\cite{van2017neural}, VQGAN~\cite{esser2021taming}, DALL-E~\cite{ramesh2021zero}, and LlamaGen~\cite{sun2024autoregressive}, may also become mainstream in the future, just like LLMs. Moreover, advanced generative models can use labels or descriptive text to conveniently control the semantics of the generated images, enabling the generated images that better fit the steganographic scenario. However, novel AR models are quite different from traditional models like PixelCNN. Is it possible to design steganography methods for existing AR models with VQ tokenizers that achieve high quality, provable security, and robustness?

In this paper, we affirm the above question. A provably secure and robust steganography based on such a semantic controllable AR image generative model, LlamaGen, is proposed. 
Considering the requirement of security and robustness, we design three modules, which are the secure message embedding module, the discrete token optimization module, and the cross-modal error-correction module.
The first module is based on the AR model, which embeds the secret message into an image token sequence in a distribution-preserving manner.
Subsequently, the token indices are decoded into an image by the image tokenizer.
The sender can utilize this module to generate high-quality secure stego images.

The receiver still faces challenges. The image tokenizer cannot accurately encode the image into correct stego tokens, and lossy social network processing exacerbates the discrepancy.
The second and third module are introduced to address these two problems. 
In the second module, an optimization process for discrete image tokens is employed to assist in the recovery of the tokens.
As for the design of a cross-modal error-correction module, with the aid of an image-to-text model, compressed error-correction information is embedded into a descriptive text about the stego image using provably secure linguistic steganography. 
The stego image is finally transmitted to the receiver along with the error-correction text, achieving provably secure robust image steganography through cross-modality error-correction.

We conducted experiments and demonstrated that our method can achieve provably secure, high-quality image robust steganography. The experimental results indicate that the proposed method significantly enhances the image quality and embedding capacity of stego images while ensuring the security and robustness of message extraction.

The main contributions of this paper are summarized:
\begin{itemize}
    \item We propose a provably secure robust image steganography method based on an auto-regressive generative model, LlamaGen, capable of generating high-quality stego images while preserving distribution.
    \item We design a robust enhancement mechanism, which includes a discrete token optimization module and a cross-modal error-correction module, to strengthen the provably secure steganography against lossy channels.
    \item Experiments verify the provable security and robustness of the proposed steganography method, and the visual effects demonstrate our significant advantage over existing methods in terms of the quality of the generated images.
\end{itemize}



\begin{figure*}[t]
    \centering
    \includegraphics[width=\textwidth]{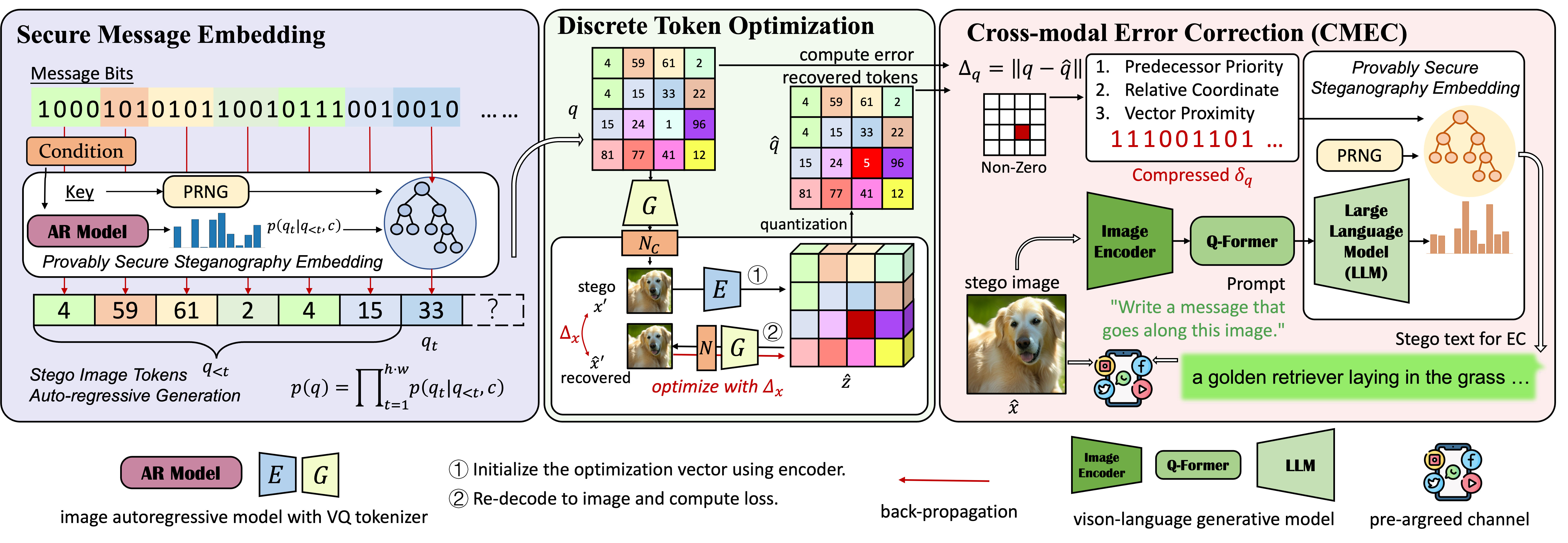}
    \caption{Overview of the proposed provably secure robust image staganography for high-quality images.  Three modules are comprised: the secure message embedding module, the discrete token optimization module, and the cross-modal error-correction module. The stego images and the stego-text are collectively transmitted to social networks to perform provably secure robust image steganography via cross-modal error-correction.}
    \label{fig:flow}
\end{figure*}

\section{Related Work}

There are two common definitions of steganographic security.
Cachin~\shortcite{cachin1998information} first proposed an information-theoretic model for steganography with passive adversaries. The adversary’s task of distinguishing between an innocent cover $c$ and a stego $s$ containing a secret message is interpreted as a ``hypothesis testing'' problem. The security of a stegosystem can be quantified by Kullback-Leibler divergence between the cover distribution $P_c$ and the stego distribution $P_s$,
\begin{equation}
    D_{KL}(P_c||P_s)=\sum_{x\in\mathcal{C}}P_c(x)\log{\frac{P_c(x)}{P_s(x)}}, 
\end{equation}
where $x$ is the object transmitted in the channel with the alphabet $\mathcal{V}$. If $D_{KL}(P_c||P_s) = 0$, the stegosystem is called \textit{perfectly secure}.
Another definition is based on computational complexity theory, proposed by Hopper \textit{et al.}~\shortcite{hopper2002provably}.
Computational security in steganography is established through a probabilistic game that distinguishes the outputs of a oracle $\mathcal{O}_{\mathcal{D}}$ that can randomly sample from the channel distribution $\mathcal{D}$ and a steganographic encoding algorithm $\textsc{Encode}_{\mathcal{D}}$. The attacker’s advantage is defined as the difference between the probability of correctly identifying the stego and the probability of incorrectly identifying a cover as a stego.
The stegosystem is called secure if all probabilistic polynomial time (PPT) adversaries $\mathcal{A}$’s advantage against the stegosystem is negligible with respect to a security parameter $\kappa$, that is:
\begin{equation}
\left|\text{Pr}\left[\mathcal{A}_\mathcal{D}^{\textsc{Encode}_{\mathcal{D}}\left(K,\cdot,\cdot\right)}=1\right]-\text{Pr}\left[\mathcal{A}_\mathcal{D}^{\mathcal{O}_{\mathcal{D}}\left(\cdot,\cdot\right)}=1\right]\right|<\text{negl}\left(\kappa\right), 
\end{equation}
where $\text{negl}\left(\kappa\right)$ is a negligible function concerning $\kappa$.


Based on the aforementioned security definitions, Hopper et al.~\shortcite{hopper2002provably} proposed a construction based on rejection sampling. Le et al.~\shortcite{van2003efficient} leveraged the duality between steganography and source coding (e.g. arithmetic coding) to encode and decode encrypted messages during the sampling process from channel distribution $\mathcal{D}$. These classic constructions needs implicit samplers or even explicit representations of $D$, which is satisfied by deep learning generative models~\cite{chen2018provably}. 
Due to the exponential time complexity of rejection sampling-based algorithms, researchers focus on implementing or improving efficient arithmetic coding-based algorithms~\cite{yang2018provably,chen2020generative,kaptchuk2021meteor}. However, their implementation inevitably distort the distribution. Zhang et al.~\shortcite{zhang2021provably} proposed ADG (adaptive dynamic grouping), grouping candidate signals with ``equal probability sums" and encoding messages using the group index. 
Ding et al.~\shortcite{ding2023discop} proposed Discop, constructing multiple ``distribution copies" during signal generation and encoding messages using the copy index, thereby avoiding distortion of the distribution. 

A suitable generative model allows these PSS constructions to be applied across various signal cover, i.e., text, audio, and images. While research into PSS for text is already well-established, its application to image cover is limited. Therefore, this paper aims to explore the potential for applying PSS to high-quality images with semantic control.


\section{Methodology}

\subsection{Approach Overview}


We propose \name{}, a novel provably secure robust image STEGanography via Cross-Modal error-correction. As shown in Figure~\ref{fig:flow}, \name{} comprises three modules: the secure message embedding module, the discrete token optimization module, and the cross-modal error correction module. 
Details of the three modules are illustrated below.

\begin{figure}[t]
    \centering
    \includegraphics[width=0.6\columnwidth]{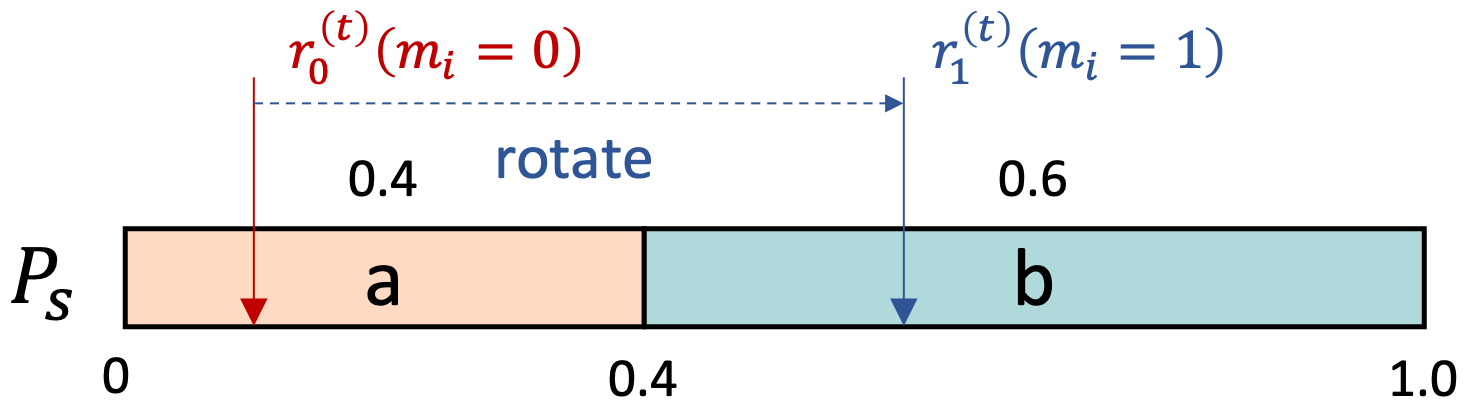}
    \caption{An example of Discop’s embedding algorithm given a distribution \{`a':[0,0.4),`b':[0.4,1.0)\}. 
    A copy of the distribution that has been shifted by $0.5$ is \{`a':[0.5,0.9),`b':[0,0.5)$\cup$[0.9,1.0)\}. 
    A random number controlled by $K$ falls into a token interval, while the number will fall into another interval after it is offset by 0.5.
    Depending on the message bit, the token into whose interval the random number falls can be selected, which is equivalent to using a copy of the distribution to represent different message bits.
}
    \label{fig:embedding}
\end{figure}

\subsection{Secure Message Steganography Module $M_1$}

This module is capable of generating a stego image with height $H$ and width $W$, 
which is deployed with the sampling process of a pre-trained AR model with a Vector Quantised (VQ) tokenizer.
The AR generative model is trained to generate a sequence of discrete image tokens $\boldsymbol{q}\in\mathbb{Q}^{h\times w}$, where $h=H/p$, $w=W/p$, $p$ is the downsample ratio of the image tokenizer, every $\boldsymbol{q}^{\left(i,j\right)}$ is a indice of a image codebook. The sequence of tokens starts from a given conditional embedding $\mathcal{H}$ and stops at the location of the pre-defined maximum length $h\cdot w$. 
Image tokens $\left(q_1, q_2, \dots, q_{h\cdot w}\right)$ are sampled by AR models in the way of next-token prediction.
Utilizing the probability distribution $p(q_t\vert q_{<t}, \mathcal{H})$ predicted by the AR model, PSS constructions such as Meteor, Discop can be deployed during the sampling phase of image token generation. At each time step $t$, the stego image token is generated as follows:
\begin{equation}
    q_t = \textsc{Encode}_{p(q_t\vert q_{<t}, \mathcal{H})}\left(K, m_t\right),
\end{equation}
where $\textsc{Encode}$ denotes the steganographic embedding algorithm used in practical deployment.
Using Discop as an example, $\textsc{Encode}$ first constructs several distribution copies based on the probability distribution, then selects the one that represents the message bits $m_t$ from the distribution copies according to the secret message to be embedded, and finally chooses the stego token $q_t$ for this time step based on the random number controlled by the steganographic key $K$. 
Figure~\ref{fig:embedding} provides a simple example of using Discop to select a token from a distribution based on a message bit.

Then, a VQ tokenizer consisting of an encoder $E$ and a decoder $G$ is used to remap the code indices $\boldsymbol{q}$ into the corresponding feature vectors $\boldsymbol{z}_{\boldsymbol{q}}$ in a discrete codebook $\mathcal{Z}$,
where $\mathcal{Z}\in \mathbb{R}^{N \times d}$ is with $N$ learnable vectors and pre-trained as well as the image tokenizer $E$ and $G$, $d$ is the dimension of $\boldsymbol{z}_{\boldsymbol{q}}$.
Then decoder $G$ converts the vectors back into image pixels  $\boldsymbol{x} \in \mathbb{R}^{H\times W \times 3}$ by:
\begin{equation}
    \boldsymbol{x}=G(\boldsymbol{z}_{\boldsymbol{q}}),
\end{equation}
where $\boldsymbol{x}$ is the generated stego image.

\subsection{Discrete Token Optimization Module $M_2$}

\begin{figure}[t]
    \centering
    \includegraphics[width=0.9\columnwidth]{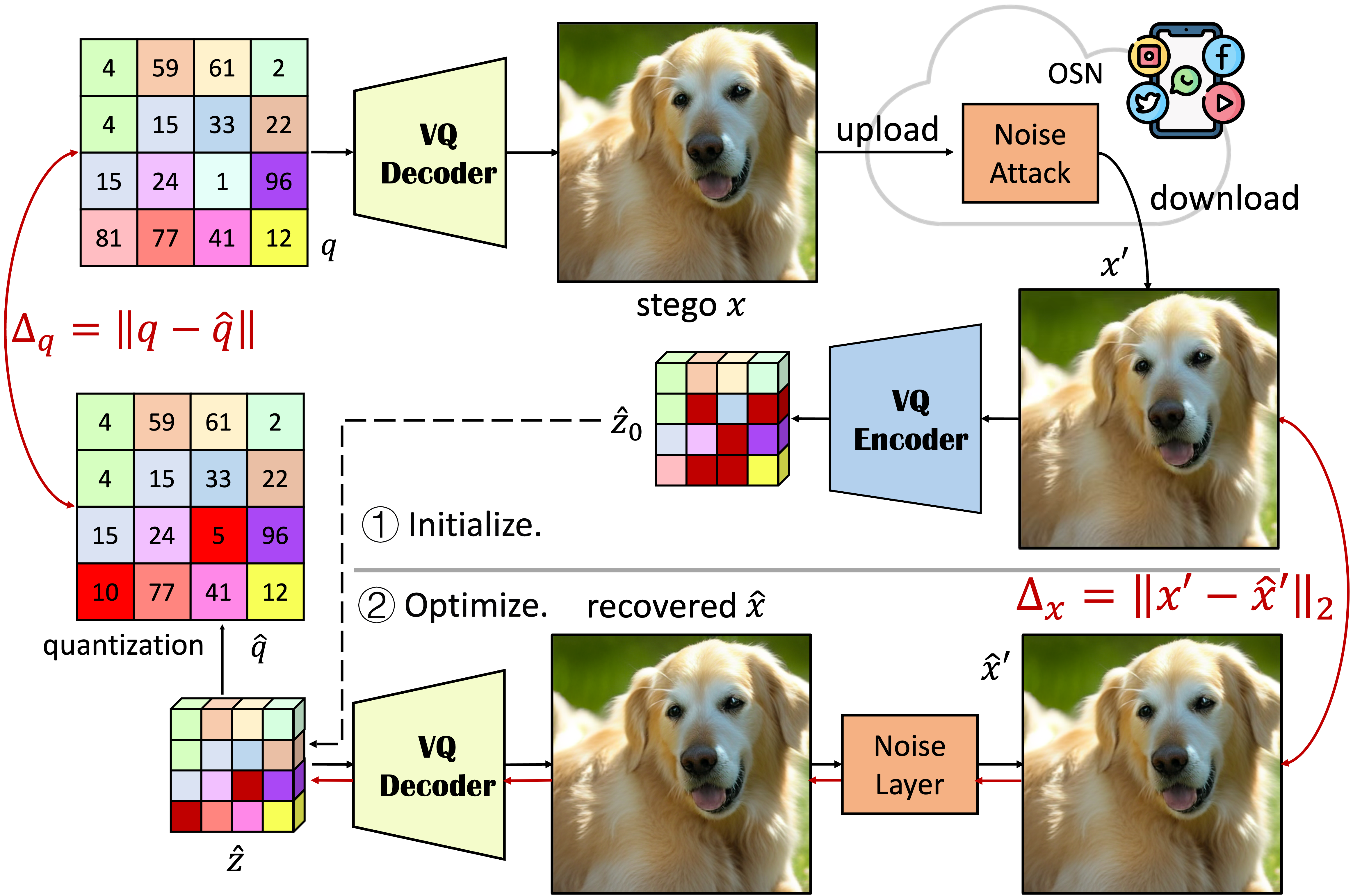}
    \caption{Flowchart of the discrete token optimization module used in the proposed provably secure and robust image steganography method.}
    \label{fig:opt}
\end{figure}

The sender then uploads the stego image to a pre-agreed communication channel with the receiver. 

Assuming the channel is lossless, the receiver can directly obtain the original generated stego image from the channel. 
However, the receiver cannot directly extract the secret message from the image pixels and must re-encode it into image tokens. 
Specifically, the encoder $E$ takes $\boldsymbol{x}$ as input and first outputs a set of continuous vectors:
\begin{equation}
\hat{\boldsymbol{z}}=E(\boldsymbol{x})\in\mathbb{R}^{h\times w\times d}.
\end{equation}
In the subsequent element-wise quantization $\mathcal{Q}(\cdot)$, each vector $\hat{\boldsymbol{z}}^{(i,j)}\in\mathbb{R}^{d}$ is quantized to its closest codebook entry $\boldsymbol{z}_{n}$:
\begin{equation}
    \hat{\boldsymbol{q}}=\mathcal{Q}(\hat{\boldsymbol{z}}):=\left(\arg\min_{\substack{n\in N}}\lVert{\hat{\boldsymbol{z}}^{(i,j)}-\boldsymbol{z}_{n}}\rVert\right),
\end{equation}
where $\hat{\boldsymbol{q}}\in\mathbb{Q}^{h\times w}$ are the indices corresponding to quantized vectors $\boldsymbol{z}_{\hat{\boldsymbol{q}}}\in\mathbb{R}^{h\times w \times d}$.

Unfortunately, the VQ tokenizers do not guarantee consistency on the vectors before and after passing through a Decoder-Encoder structure.
Formally, the loss between stego tokens $\boldsymbol{q}$ and re-encoded tokens $\hat{\boldsymbol{q}}$ can be denoted as:
\begin{equation}
    \Delta_{\boldsymbol{q}}=\lVert \boldsymbol{q}-\hat{\boldsymbol{q}}\rVert.
\end{equation}


To relatively accurately extract the secret message, it is necessary to make $\Delta_{\boldsymbol{q}}$ as small as possible. We use the reversed tokens $\hat{\boldsymbol{q}}$ to regenerate a recovered image for optimization.
A differentiable noise layer $\mathcal{N}$ designed to simulate the noise attack $\mathcal{N}_C$ of the channel $C$ is introduced, ensuring that the recovered image undergoes lossy operations similar to those experienced by the stego image.
Then the receiver can get a lossy recovered image:
\begin{equation}
    \hat{\boldsymbol{x}}'=\mathcal{N}(\hat{\boldsymbol{x}})=\mathcal{N}\left(G\left(\boldsymbol{z}_{\hat{\boldsymbol{q}}}\right)\right).
\end{equation}
The difference between the lossy recovered image $\hat{\boldsymbol{x}}'$ and the lossy stego image $\boldsymbol{x}'$ can be denoted as:
\begin{align}
    \Delta_{\boldsymbol{x}} &= \lVert \boldsymbol{x}'-\hat{\boldsymbol{x}}'\rVert_2 \\
    &= \lVert \mathcal{N}_C(G(\boldsymbol{z}_{\boldsymbol{q}}))-\mathcal{N}\left(G\left(\boldsymbol{z}_{\hat{\boldsymbol{q}}}\right)\right)\rVert_2,
\end{align}
If $\hat{\boldsymbol{q}}$ is identical to $\boldsymbol{q}$, then $\boldsymbol{z}_{\hat{\boldsymbol{q}}}$ is the same as $\boldsymbol{z}_{\boldsymbol{q}}$, and $\Delta_{\boldsymbol{x}}$ will be quite small. 
Let $\Delta_{\boldsymbol{x}}$ be a loss function of $\boldsymbol{q}$, and an optimization method can be used to obtain a $\hat{\boldsymbol{q}}$ that is as close as possible to $\boldsymbol{q}$.

Since the tokens are discrete integers, there is no gradient at $\boldsymbol{q}$ and $\hat{\boldsymbol{q}}$. For generation images of larger sizes, the discrete optimization of the \( h\times w\times N \) dimensions is quite challenging. Therefore, we use the differentiable continuous vectors $\hat{\boldsymbol{z}}\in\mathbb{R}^{h\times w\times d}$ to replace $\hat{\boldsymbol{q}}$ for optimization based on gradient descent:
\begin{equation}
    \hat{\boldsymbol{z}}\leftarrow \left[\hat{\boldsymbol{z}}-\gamma_{\hat{\boldsymbol{z}}}\frac{\partial \Delta_{\boldsymbol{x}}}{\partial \hat{\boldsymbol{z}}}\right],
\end{equation}
where $\gamma_{\hat{\boldsymbol{z}}}$ denotes the learning rate of gradient descent.

Ultimately, after the optimization process, we convert \( \hat{\boldsymbol{z}} \) back into discrete tokens $\hat{\boldsymbol{q}}=\mathcal{Q}(\hat{\boldsymbol{z}})$, which is more similar to $\boldsymbol{q}$ than re-encode $\boldsymbol{x}'$ directly. 
The whole discrete token optimization module is shown in Figure~\ref{fig:opt}.


\subsection{Cross-Modal Error-Correction Module $M_3$}

Upon observation, we found that even if $\Delta_{\boldsymbol{x}}$ converges to a considerably low level during the optimization process, there are still some recovered tokens that differ from the original stego tokens.
We introduce additional error-correction mechanisms to enhance the robustness of steganography in this module.


Once the sender uploads the generated stego image $\boldsymbol{x}$ to the selected channel, the sender has the ability to carry out a complete discrete token optimization process, just as the receiver would do during extraction. 
If there remains some tokens that cannot be recovered, the sender can supplement this part to the receiver in some other way.
Specifically, let $\boldsymbol{\delta}_{\boldsymbol{q}}$ be a set that represents the non-zero elements from $\Delta_{\boldsymbol{q}}$,
\begin{equation}
    \boldsymbol{\delta}_{\boldsymbol{q}} = \{\left((i,j),\boldsymbol{q}^{(i,j)}\right)\mid{\Delta_{\boldsymbol{q}}}^{(i,j)} \neq 0 \}.
\end{equation}
The error-correction module embeds \( \boldsymbol{\delta}_{\boldsymbol{q}} \) as a secret message into a piece of text using a PSS method based on generative models. The stego text is then conveyed to the receiver.
To further strengthen the semantic connection between the stego text used for error correction and the original stego image, we opt to utilize a pre-trained vision-language model, which consists of a frozen image encoder $E_B$, a pre-trained querying transformer (Q-Former) $QF_B$ used for bridging the modality gap, and a large language model $LLM$ for generation.
Firstly, the image encoder $E_B$ and the Q-Former $QF_B$ jointly take responsibility for extracting the lossy stego image $\boldsymbol{x}'$ that has been processed by the channel into a visual representation $\mathcal{H}_{\boldsymbol{x}'}$ that can be understood by the $LLM$, which can be denoted as:
\begin{equation}
    \mathcal{H}_{\boldsymbol{x}'} = {QF}_B(E_B({\boldsymbol{x}'}),\mathcal{H}_t),
\end{equation}
where $\mathcal{H}_t$ represents the instruction text or question that can be input during the Q-Former encoding process.
Then the $LLM$ generates a corresponding descriptive text for $\boldsymbol{x}'$ with $\mathcal{H}_{\boldsymbol{x}'}$ as the context, while steganographic methods like Discop are employed to embed $\boldsymbol{\delta}_{\boldsymbol{q}}$ within it.
Due to the limited carrying capacity of text, to ensure complete error correction as much as possible, we also need to compress \( \boldsymbol{\delta}_{\boldsymbol{q}} \).
To send as little additional information as possible while achieving the strongest robustness, three principles are adhered to when embedding error-correction information, namely:
\subsubsection{Predecessor Priority.}
Errors that appear early in the image token sequence can affect subsequent tokens, necessitating the prioritization of error correction for preceding ones.
\subsubsection{Relative Coordinate.}
Most token reconstruction errors tend to cluster. Except for the first token, we represent the occurrence location of each erroneous token using relative coordinates $\delta_1$ from the position where the previous erroneous token appeared.
To reduce the volume of error correction information, we set a maximum relative coordinate threshold $\lambda_1$. Tokens exceeding the maximum relative coordinate will not be corrected.

\subsubsection{Vector Proximity.}
During the generation of image tokens, the sampling is restricted to the top-$k$ tokens. By calculating the distance between all the top-$k$ vectors corresponding to samplable tokens and the vector corresponding to the incorrect reconstructed token after optimization, only the sorted sequence numbers $\delta_2$ corresponding to the correct tokens are transmitted during error correction.
Given a set of vectors $\boldsymbol{z} = \{\boldsymbol{z}_1, \boldsymbol{z}_2, \dots, \boldsymbol{z}_k\}$ after an optimization process, where each $\boldsymbol{z}_i$ corresponds to a samplable top-$k$ token $q_i$. Let $\boldsymbol{z}_e$ be the vector corresponding to the incorrectly reconstructed token. We calculate the distance between each $\boldsymbol{z}_i$ and $\boldsymbol{z}_e$, $\Delta_{\boldsymbol{z},i} = \|\boldsymbol{z}_i - \boldsymbol{z}_e\|,$ $i = 1, 2, \dots, k.$
The distances $\Delta_{\boldsymbol{z},1}, \Delta_{\boldsymbol{z},2}, \dots, \Delta_{\boldsymbol{z},k}$ are then sorted, and the corresponding indices are $o_1, o_2, \dots, o_k$. During error correction, only the sequence number $o_j$ corresponding to the correct token is transmitted, that is $\delta_2 = o_j$ where $\boldsymbol{z}_j = \boldsymbol{z}_{q}.$
Similar to the maximum relative coordinate value, a maximum relative sequence threshold $\lambda_2$ will also be set; tokens exceeding this will not be corrected.

The compressed error-correction $\boldsymbol{\delta}_{\boldsymbol{q}}$ can be denoted as:
\begin{equation}
    \boldsymbol{\delta}_{\boldsymbol{q}} = \{\left(\left(\delta_1,\delta_2\right)^{(i,j)}\right)\mid 1\leq i \leq h, 1\leq j \leq w,{\Delta_{\boldsymbol{q}}}^{(i,j)} \neq 0 \},
\end{equation}
where $0\leq \delta_1<2^{\lambda_1}$, $0\leq \delta_2<2^{\lambda_2}$.
Every $\left(\delta_1,\delta_2\right)$ is encoded into binary numbers and encrypted, waiting for steganographic embedding.
At each time step $t$ of sampling process of the vision-language model, the stego text token $l_t$ for error-correction is generated as follows:
\begin{equation}
    l_t = \textsc{Encode}_{p(l_t\vert l_{<t}, \mathcal{H}_{\boldsymbol{x}'})}\left(K, \delta_t\right).
\end{equation}
Assuming $\textsc{Encode}$ has an embedding rate of $\rho$ bits per token on the LLM, then the number $\tau$ of erroneous image tokens that the stego text tokens of length $\ell$ can correct can be calculated as:
\begin{equation}
    \tau = \left\lfloor 1+\frac{\rho\cdot \ell-\lfloor\log_2(h\cdot w)\rfloor+\lambda_2}{\lambda_1 + \lambda_2} \right\rfloor.
\end{equation}

After the steganographic process is completed, the stego text corresponding to stego text tokens $\boldsymbol{l}=\left(l_1,l_2,\dots,l_{\ell}\right)$, along with the stego image $\boldsymbol{x}$, is transmitted to the receiver. 
The receiver can then extract error-correction information from the stego text to assist in message extraction from the stego image. 
Ultimately, robust provably secure image steganography is achieved.


\subsection{Complexity}

The time complexity of our method can be evaluated in three modules. In $M_1$, the time to generate the stego image includes the predicting time of the token distribution, the embedding time of secret message, and the generating time of the image from the tokens. The embedding time depends on the algorithm used. The complexity of optimizing in $M_2$ is $O(T(3\cdot H\cdot W+d)$, where $T$ is the numbers of iterations, $d$ is the dimension of vector. $M_3$'s time includes the time to compute error correction information and to generate the stego text. The first time is related to the number of tokens, the top-$k$ value, and the dimension of the vectors. The time complexity is $O(h\cdot w(k\cdot d+k\log k))$.

\subsection{Proof of Security}

In our method, both the stego image with embedded secret message and the stego text with embedded error correction information are transmitted through public channels, and security needs to be guaranteed at the same time. For stego text, since the security of the embedding algorithm used has been proven, in this paper we only discuss the security of the image steganographic embedding algorithm, that is, the undetectability of the stego image from the normal generated image.

\begin{figure*}[t]
    \centering
    \includegraphics[width=\textwidth]{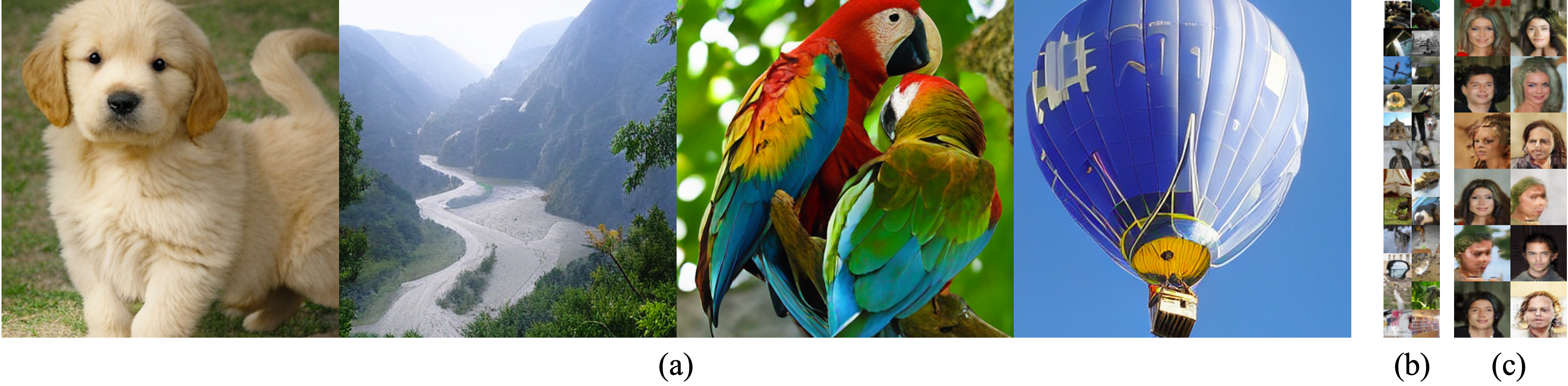}
    \caption{Visual results of generated stego images. All images are scaled to a suitable display size at the same ratio. (a) Ours; (b) Discop-ImageGPT~\cite{ding2023discop}; (c) PARIS~\cite{yang2023provably}.}
    \label{fig:show}
\end{figure*}

Assume that a PPT adversary $\mathcal{A}$ possesses a non-negligible advantage to distinguish the generated stego image $\boldsymbol{x}$ from a randomly sampled cover image $\boldsymbol{x}_c$ by the same model, which can be defined as:
\begin{equation}
    \left|\text{Pr}[\mathcal{A}(\boldsymbol{x})=1]-\text{Pr}[\mathcal{A}(\boldsymbol{x}_c)=1]\right|=\epsilon,
\end{equation}
where $\epsilon$ denotes a non-negligible quantity relative to the length of shared key $K$, indicating that $\mathcal{A}$ is able to distinguish between $\boldsymbol{x}_c$ and $\boldsymbol{x}$.
In this paper, an image is generated by $\boldsymbol{x}=G(\boldsymbol{z}_{\boldsymbol{q}})$. We denote the sequence of tokens used to generate the cover image as $\boldsymbol{q}_c$ and the tokens used to generate the stego image as $\boldsymbol{q}$. Hence, the advantage of $\mathcal{A}$ can be calculated as:
\begin{equation}
    \left|\text{Pr}[\mathcal{A}(G(\boldsymbol{z}_{\boldsymbol{q}}))=1]-\text{Pr}[\mathcal{A}(G(\boldsymbol{z}_{\boldsymbol{q}_{c}}))=1]\right|=\epsilon.
\end{equation}
That is, $\mathcal{A}$'s advantage to distinguish between $\boldsymbol{x}$ and $\boldsymbol{x}_c$ can be reduced to an advantage to distinguish between $\boldsymbol{q}$ and $\boldsymbol{q}_c$. For each time step $t$, $q_t$ is obtained by the steganographic embedding algorithm \textsc{Encode} based on shared key $K$ and message bits $m_t$, while $q_{c,t}$ is determined by a random sampling algorithm \textsc{Sample} with a random number $r_t$. Hence, the advantage is:
\begin{align}
    |\text{Pr}[\mathcal{A}(\textsc{Encode}_{p(q_t\vert q_{<t}, \mathcal{H})}\left(K, m_t\right))&=1]\\\nonumber
    -\text{Pr}[\mathcal{A}(\textsc{Sample}_{p(q_t\vert q_{<t}, \mathcal{H})}\left(r_t\right))&=1]|=\epsilon.
\end{align}\label{eq:prove}

Based on the previously proposed PSS constructions~\cite{hopper2002provably,kaptchuk2021meteor,de2022perfectly,ding2023discop}, the aforementioned advantages can be reduced to $\mathcal{A}$'s ability to distinguish between a uniformly distributed random number obtained by encrypting with an encryption algorithm and a random number directly sampled from a uniform distribution in polynomial time.
However, during steganography, a computationally secure symmetric encryption scheme is utilized. Therefore, the non-negligible advantage cannot hold, indicating that the cover and the stego are indistinguishable in polynomial time, validating the computational security of the proposed image steganography method. Q.E.D.

\section{Experiments}

In this section, we conduct experiments to present the performance of \name{} mainly in terms of visualization and robustness, and compare \name{} with previous provably secure image steganography methods. The platform is Pytorch 2.3.1 and NVIDIA A6000.

\subsubsection{Secure Message Steganography Module $\text{M}_1$}
In our experiments, we utilize a VQGAN with a downsampling rate of $16$ as the VQ tokenizer. The codebook vector dimension is $8$, codebook size is $16384$. We employ an AR model with $3$ billion parameters based on the Llama architecture for generating image tokens. The training of both VQ tokenizer and AR model is on ImageNet train set, using the resolution of $256\times256$ and random crop data augmentation.
Top-$k$ is set to $2000$.
In the steganography experiments, we directly use the pre-trained models for generation without retraining them.
The generated image size is set to $384\times 384$.
The category labels used for generating cover and stego images are also sourced from ImageNet. 
Therefore, each cover or stego image corresponds to a token sequence of length $576$.

\subsubsection{Discrete Token Optimization Module $\text{M}_2$}

When simulating a lossy channel with a noise layer, 
JPEG-SS is used to simulate JPEG noise in a differentiable manner since it performs better than JPEG-MASK according to Yang et al.~\shortcite{yang2023provably}.
The momentum-based optimizer Adam is adopted with an initial learning rate of $0.002$.
The number of optimization steps is set to $10,000$.


\subsubsection{Cross-Modality Error Correction Module $\text{M}_3$}
As for the image-to-text model, a InstructBLIP~\cite{instructblip} model with a 7 billion parameter Vicuna language model is used.
$\lambda_1$ is set to $8$, as is $\lambda_2$. Max token length is set to $200$.

\begin{table}[t]
\centering
\setlength{\tabcolsep}{1mm}
\begin{tabular}{@{}ccccc@{}}
\toprule
Method & Model    & Semantic & Robust & Resolution  \\ \midrule
Discop-ImageGPT & AR & weak & weak & $32\times32$  \\
PARIS  & GAN    & weak & strong &$64\times64$  \\
StegaStyleGAN & GAN & weak & strong &$256\times256$ \\
\name{}   & AR & strong & strong &$384\times384$ \\ 

\bottomrule
\end{tabular}
\caption{Comparison of resolution of generated images with other provably secure steganography methods.}\label{tab:resolution}
\end{table}

\subsection{Experimental Results}

\subsubsection{Visual Quality} 

\begin{figure}[t]
    \centering
    \includegraphics[width=0.8\columnwidth]{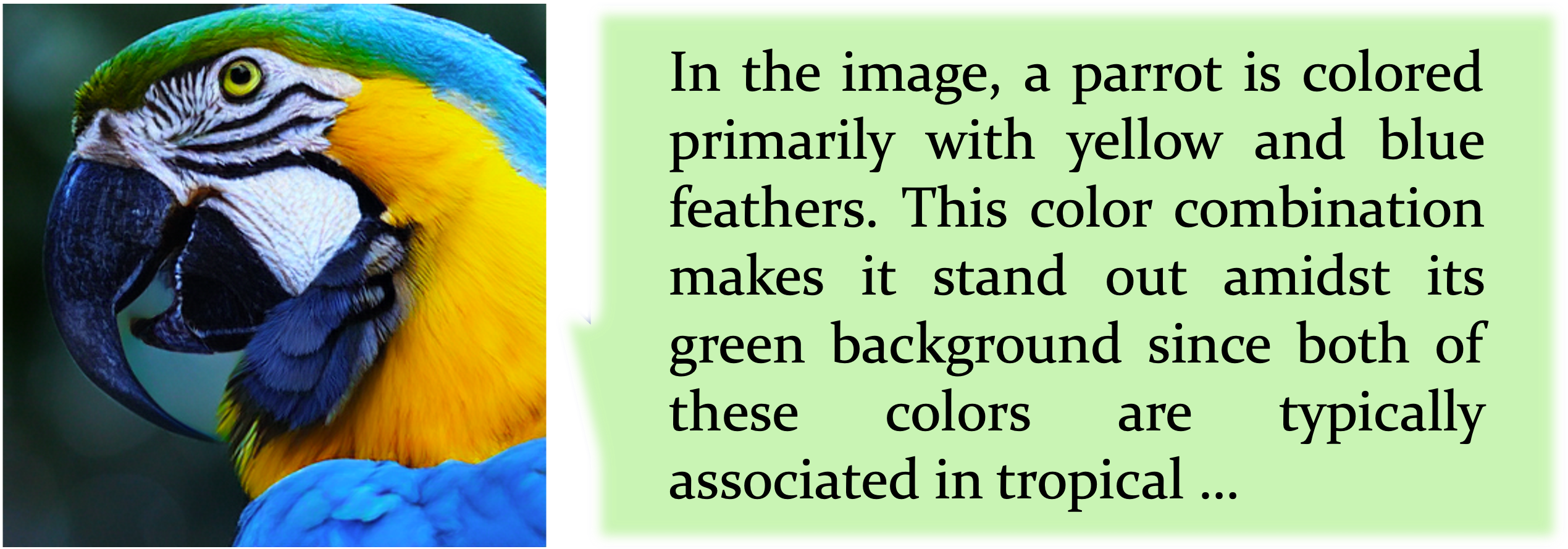}
    \caption{Example of stego image and its corresponding semantically consistent stego text for error correction.}\label{fig:cross-modal}
\end{figure}

We focus on two aspects of visual quality: one is the comparison of the quality of stego images that different steganographic methods can generate, and the other is the comparison of quality between randomly sampled cover images and stego images generated by the provably secure and robust steganographic method. For the first aspect, we are mainly concerned with the resolution of the images.
Table~\ref{tab:resolution} presents a comparison of the resolutions of the stego images that \name{} and other methods can generate. As illustrated in Figure~\ref{fig:show}, our \name{} can generate stego images with higher resolution, greater diversity, and better visual quality.
Figure~\ref{fig:cross-modal} shows the stego image and its corresponding error-correcting text, both of which have consistent semantics.


\begin{table}[t]
\centering
\setlength{\tabcolsep}{1mm}
\begin{tabular}{@{}cccccccc@{}}
\toprule
\multicolumn{2}{c}{\multirow{2}{*}{Noise}} & \multicolumn{2}{c}{$\text{w/ }\text{M}_1$} & \multicolumn{2}{c}{$\text{w/ } \text{M}_{1,2}$} & \multicolumn{2}{c}{$\text{w/ } \text{M}_{1,2,3}$} \\ \cmidrule(l){3-8} 
\multicolumn{2}{c}{} & $R_q$ & $Cap$ & $R_q$ & $Cap$ & $R_q$ & $Cap$ \\ \midrule
\multicolumn{2}{c}{$-$} & 91.89 & 106 & 99.89 & 3803 & 99.98 & 4285 \\ \midrule
\multirow{3}{*}{JPEG} & QF 95 & 89.56 & 66 & 99.30 & 2793 & 99.81 & 3936 \\
 & QF 85 & 84.60 & 47 & 98.28 & 2551 & 99.34 & 3861 \\
 & QF 75 & 78.65 & 51 & 97.39 & 1935 & 98.42 & 3293 \\ \midrule
G.N. & 0.01 & 32.88 & 15 & 93.33 & 1699 & 95.31 & 2888 \\ \midrule
\multirow{2}{*}{Scale} & $0.5\times$ & 68.29 & 34 & 99.75 & 3725 & 99.99 & 4174 \\
 & $2.0\times$ & 89.73 & 82 & 99.84 & 3809 & 100.0 & 4292 \\ \bottomrule
\end{tabular}
\caption{Performance of the proposed \name{} against JPEG compression and other noise of different strengths.}\label{tab:robust}
\end{table}

\subsubsection{Robustness}

We evaluate the robustness mainly using token recovery rate $R_q$.
Effective capacity ($Cap$) is calculated as the maximum number of message bits that can be successfully decoded and extracted with the error-correcting capability of the system before encountering an error that exceeds the system's correction threshold.
We generate $50$ stego images using \name{} with random message and attempt to extract the message from it.
Specifically, we extract the message immediately after re-encoding the stego images, after the optimization process, and following the error correction. The results of these three extractions are recorded as $\text{w/ }\text{M}_1$, $\text{w/ } \text{M}_{1,2}$, and $\text{w/ } \text{M}_{1,2,3}$, respectively, as shown in Table~\ref{tab:robust}.
It can be seen that \name{} can almost achieve lossless embedding and extraction of high-capacity messages over a lossless channel after passing through all three modules. 
It is also worth noting that due to the characteristics of AR models, preceding errors will affect the extraction of subsequent messages. Therefore, $R_q$ is not entirely proportional to the effective capacity.

\subsubsection{Security}

To verify the security of the algorithm, three deep-learning-based steganalyzers, namely ConvNet~\cite{covnet}, SRNet~\cite{boroumand2018deep}, and LWENet~\cite{weng2022lightweight}, are employed to distinguish the cover image and stego image.
In the experiment, the detection error rate $\Bar{P}_E=\frac{P_{FA}+P_{MD}}{2}$ is tested respectively, where $P_{FA}$ denotes the false alarm rate and $P_{MD}$ denotes the missed detection rate.
For training three steganalyzers, $4000$ randomly sampled generated images are selected as covers, and $4000$ secret message-driven generated images are used as stegos. $1000$ covers and $1000$ stegos are used for the ﬁnal test. The experimental results are shown in Table~\ref{tab:steganalysis}. Remarkably, the detection error rates remain closely to $0.5$, indicating that the cover and stego images are indistinguishable. Our comparison is also within provably secure steganographic methods, which similarly provide theoretical guarantees; as shown in their papers~\cite{ding2023discop,yang2023provably,su2024stegastylegan}, their detection error rates are also around $0.5$. Experimental results show that our proposed method, like previous secure image steganography methods, can resist detection by existing steganalyzers.

\begin{table}[t]
\centering
\begin{tabular}{@{}cccc@{}}
\toprule
Steganalyzer & ConvNet & LWENet & SRNet     \\ \midrule
$\Bar{P}_E$  & 0.5014 & 0.5043 & 0.4980    \\\bottomrule
\end{tabular}
\caption{Detection error rate $\Bar{P}_E$ against different steganalyzers.}\label{tab:steganalysis}
\end{table}


\begin{table}[t]
\centering
\begin{tabular}{@{}ccccc@{}}
\toprule
$\ell$ (token) & 50 & 100 & 200 & 500     \\ \midrule
Payload (bit)  & 75 & 217 & 628 & 2037   \\
$R_q$ ($\%$)   & 99.54 & 99.67 & 99.81 & 99.80   \\
$Cap$ (bit)   & 3608 & 3903 & 3935 & 3986   \\\bottomrule
\end{tabular}
\caption{Robustness of \name{} under JPEG Compression (QF=95) with different max token lengths.}\label{tab:textlen}
\end{table}



\subsubsection{Effectiveness of Error Correction}

Table~\ref{tab:textlen} shows the number of bits that can be embedded in the text, the reconstruction accuracy of image tokens after text error correction, and the effective capacity of secret messages, all varying with the maximum text token length. It can be observed that as the text length increases, the number of bits that can be embedded in the text increases significantly. We believe this is because, with the increase in text sequence length, the constraints imposed by the image and the initial prompt on text generation become weaker. Increasing the length of the error-correcting stego text can enhance the robustness of \name{}, which aligns with our expectations. 


\section{Conclusion}

In this paper, we propose \name{}, for the first time, achieving provably secure and robust image steganography on AR image generation models with VQ tokenizer. \name{} comprises three modules. The first secure message embedding module embeds secret message into stego images without altering any distribution. The second discrete token optimization module helps to recover the lost stego tokens during re-encoding and the lossy channel. The third cross-modal error-correction module utilizes an image-to-text model to generate semantically consistent stego text corresponding to the stego image with error-correction message embedded in it. Experiments on LlamaGen demonstrate that \name{} can generate high-quality stego images. We have provided theoretical proofs for the security of the proposed image steganography method and experimental validation against steganalyzers. The designed cross-modality error-correction module effectively enhances the robustness of steganography, ensuring that the method can extract secret messages with a high payload under various types of noise.


\section{Acknowledgments}
This work was supported in part by the National Natural Science Foundation of China under Grant 62472398, U2336206, 62402469, and 62121002.
\bibliography{main}

\begin{thebibliography}{38}
\providecommand{\natexlab}[1]{#1}

\bibitem[{Achiam et~al.(2023)Achiam, Adler, Agarwal, Ahmad, Akkaya, Aleman,
  Almeida, Altenschmidt, Altman, Anadkat et~al.}]{achiam2023gpt}
Achiam, J.; Adler, S.; Agarwal, S.; Ahmad, L.; Akkaya, I.; Aleman, F.~L.;
  Almeida, D.; Altenschmidt, J.; Altman, S.; Anadkat, S.; et~al. 2023.
\newblock {GPT-4} technical report.
\newblock \emph{arXiv preprint arXiv:2303.08774}.

\bibitem[{Almeida et~al.(2024)Almeida, Nunes, Engelmann, Wiegmann, and
  de~Ara{\'u}jo}]{almeida2024exploring}
Almeida, G.~F.; Nunes, J.~L.; Engelmann, N.; Wiegmann, A.; and de~Ara{\'u}jo,
  M. 2024.
\newblock Exploring the psychology of LLMs’ moral and legal reasoning.
\newblock \emph{Artificial Intelligence}, 333: 104145.

\bibitem[{Boroumand, Chen, and Fridrich(2018)}]{boroumand2018deep}
Boroumand, M.; Chen, M.; and Fridrich, J. 2018.
\newblock Deep residual network for steganalysis of digital images.
\newblock \emph{IEEE Transactions on Information Forensics and Security},
  14(5): 1181--1193.

\bibitem[{Cachin(1998)}]{cachin1998information}
Cachin, C. 1998.
\newblock An information-theoretic model for steganography.
\newblock In \emph{International Workshop on Information Hiding}, 306--318.
  Springer.

\bibitem[{Cachin et~al.(2005)}]{cachin2005digital}
Cachin, C.; et~al. 2005.
\newblock Digital Steganography.

\bibitem[{Chen et~al.(2018)Chen, Zhou, Zhao, Chen, Zhang, and
  Yu}]{chen2018provably}
Chen, K.; Zhou, H.; Zhao, H.; Chen, D.; Zhang, W.; and Yu, N. 2018.
\newblock When provably secure steganography meets generative models.
\newblock \emph{arXiv preprint arXiv:1811.03732}.

\bibitem[{Chen et~al.(2020)Chen, Radford, Child, Wu, Jun, Luan, and
  Sutskever}]{chen2020generative}
Chen, M.; Radford, A.; Child, R.; Wu, J.; Jun, H.; Luan, D.; and Sutskever, I.
  2020.
\newblock Generative pretraining from pixels.
\newblock In \emph{International conference on machine learning}, 1691--1703.
  PMLR.

\bibitem[{Dai et~al.(2023)Dai, Li, Li, Tiong, Zhao, Wang, Li, Fung, and
  Hoi}]{instructblip}
Dai, W.; Li, J.; Li, D.; Tiong, A. M.~H.; Zhao, J.; Wang, W.; Li, B.; Fung, P.;
  and Hoi, S. 2023.
\newblock InstructBLIP: Towards General-purpose Vision-Language Models with
  Instruction Tuning.
\newblock arXiv:2305.06500.

\bibitem[{de~Witt et~al.(2022)de~Witt, Sokota, Kolter, Foerster, and
  Strohmeier}]{de2022perfectly}
de~Witt, C.~S.; Sokota, S.; Kolter, J.~Z.; Foerster, J.~N.; and Strohmeier, M.
  2022.
\newblock Perfectly Secure Steganography Using Minimum Entropy Coupling.
\newblock In \emph{The Eleventh International Conference on Learning
  Representations}.

\bibitem[{Deng et~al.(2019)Deng, Chen, Luo, and Luo}]{covnet}
Deng, X.; Chen, B.; Luo, W.; and Luo, D. 2019.
\newblock Fast and Effective Global Covariance Pooling Network for Image
  Steganalysis.
\newblock \emph{Proceedings of the ACM Workshop on Information Hiding and
  Multimedia Security}.

\bibitem[{Ding et~al.(2023)Ding, Chen, Wang, Zhao, Zhang, and
  Yu}]{ding2023discop}
Ding, J.; Chen, K.; Wang, Y.; Zhao, N.; Zhang, W.; and Yu, N. 2023.
\newblock Discop: Provably secure steganography in practice based on"
  distribution copies".
\newblock In \emph{2023 IEEE Symposium on Security and Privacy (SP)},
  2238--2255. IEEE.

\bibitem[{Du et~al.(2024)Du, Chang, Hospedales, Song, and
  Ma}]{du2024demofusion}
Du, R.; Chang, D.; Hospedales, T.; Song, Y.-Z.; and Ma, Z. 2024.
\newblock Demofusion: Democratising high-resolution image generation with no
  \$\$\$.
\newblock In \emph{Proceedings of the IEEE/CVF Conference on Computer Vision
  and Pattern Recognition}, 6159--6168.

\bibitem[{Esser, Rombach, and Ommer(2021)}]{esser2021taming}
Esser, P.; Rombach, R.; and Ommer, B. 2021.
\newblock Taming transformers for high-resolution image synthesis.
\newblock In \emph{Proceedings of the IEEE/CVF conference on computer vision
  and pattern recognition}, 12873--12883.

\bibitem[{Ge et~al.(2024)Ge, Hua, Mei, Tan, Xu, Li, Zhang
  et~al.}]{ge2024openagi}
Ge, Y.; Hua, W.; Mei, K.; Tan, J.; Xu, S.; Li, Z.; Zhang, Y.; et~al. 2024.
\newblock Openagi: When llm meets domain experts.
\newblock \emph{Advances in Neural Information Processing Systems}, 36.

\bibitem[{Goodfellow et~al.(2020)Goodfellow, Pouget-Abadie, Mirza, Xu,
  Warde-Farley, Ozair, Courville, and Bengio}]{goodfellow2020generative}
Goodfellow, I.; Pouget-Abadie, J.; Mirza, M.; Xu, B.; Warde-Farley, D.; Ozair,
  S.; Courville, A.; and Bengio, Y. 2020.
\newblock Generative adversarial networks.
\newblock \emph{Communications of the ACM}, 63(11): 139--144.

\bibitem[{Hopper, Langford, and Von~Ahn(2002)}]{hopper2002provably}
Hopper, N.~J.; Langford, J.; and Von~Ahn, L. 2002.
\newblock Provably secure steganography.
\newblock In \emph{Advances in Cryptology—CRYPTO 2002: 22nd Annual
  International Cryptology Conference Santa Barbara, California, USA, August
  18--22, 2002 Proceedings 22}, 77--92. Springer.

\bibitem[{Kaptchuk et~al.(2021)Kaptchuk, Jois, Green, and
  Rubin}]{kaptchuk2021meteor}
Kaptchuk, G.; Jois, T.~M.; Green, M.; and Rubin, A.~D. 2021.
\newblock Meteor: Cryptographically secure steganography for realistic
  distributions.
\newblock In \emph{Proceedings of the 2021 ACM SIGSAC Conference on Computer
  and Communications Security}, 1529--1548.

\bibitem[{Karras, Laine, and Aila(2019)}]{karras2019style}
Karras, T.; Laine, S.; and Aila, T. 2019.
\newblock A style-based generator architecture for generative adversarial
  networks.
\newblock In \emph{Proceedings of the IEEE/CVF conference on computer vision
  and pattern recognition}, 4401--4410.

\bibitem[{Radford et~al.(2019)Radford, Wu, Child, Luan, Amodei, Sutskever
  et~al.}]{radford2019language}
Radford, A.; Wu, J.; Child, R.; Luan, D.; Amodei, D.; Sutskever, I.; et~al.
  2019.
\newblock Language models are unsupervised multitask learners.
\newblock \emph{OpenAI blog}, 1(8): 9.

\bibitem[{Ramesh et~al.(2021)Ramesh, Pavlov, Goh, Gray, Voss, Radford, Chen,
  and Sutskever}]{ramesh2021zero}
Ramesh, A.; Pavlov, M.; Goh, G.; Gray, S.; Voss, C.; Radford, A.; Chen, M.; and
  Sutskever, I. 2021.
\newblock Zero-shot text-to-image generation.
\newblock In \emph{International conference on machine learning}, 8821--8831.
  Pmlr.

\bibitem[{Sedighi, Cogranne, and Fridrich(2015)}]{sedighi2015content}
Sedighi, V.; Cogranne, R.; and Fridrich, J. 2015.
\newblock Content-adaptive steganography by minimizing statistical
  detectability.
\newblock \emph{IEEE Transactions on Information Forensics and Security},
  11(2): 221--234.

\bibitem[{Song and Ermon(2019)}]{song2019generative}
Song, Y.; and Ermon, S. 2019.
\newblock Generative modeling by estimating gradients of the data distribution.
\newblock \emph{Advances in neural information processing systems}, 32.

\bibitem[{Su, Ni, and Sun(2024)}]{su2024stegastylegan}
Su, W.; Ni, J.; and Sun, Y. 2024.
\newblock StegaStyleGAN: Towards Generic and Practical Generative Image
  Steganography.
\newblock In \emph{Proceedings of the AAAI Conference on Artificial
  Intelligence}, volume~38, 240--248.

\bibitem[{Sun et~al.(2024)Sun, Jiang, Chen, Zhang, Peng, Luo, and
  Yuan}]{sun2024autoregressive}
Sun, P.; Jiang, Y.; Chen, S.; Zhang, S.; Peng, B.; Luo, P.; and Yuan, Z. 2024.
\newblock Autoregressive Model Beats Diffusion: Llama for Scalable Image
  Generation.
\newblock \emph{arXiv preprint arXiv:2406.06525}.

\bibitem[{Tulsiani and Gupta(2021)}]{tulsiani2021pixeltransformer}
Tulsiani, S.; and Gupta, A. 2021.
\newblock PixelTransformer: Sample Conditioned Signal Generation.
\newblock In \emph{International Conference on Machine Learning}, 10455--10464.
  PMLR.

\bibitem[{Van~den Oord et~al.(2016)Van~den Oord, Kalchbrenner, Espeholt,
  Vinyals, Graves et~al.}]{van2016conditional}
Van~den Oord, A.; Kalchbrenner, N.; Espeholt, L.; Vinyals, O.; Graves, A.;
  et~al. 2016.
\newblock Conditional image generation with pixelcnn decoders.
\newblock \emph{Advances in neural information processing systems}, 29.

\bibitem[{Van Den~Oord, Kalchbrenner, and Kavukcuoglu(2016)}]{van2016pixel}
Van Den~Oord, A.; Kalchbrenner, N.; and Kavukcuoglu, K. 2016.
\newblock Pixel recurrent neural networks.
\newblock In \emph{International conference on machine learning}, 1747--1756.
  PMLR.

\bibitem[{Van Den~Oord, Vinyals et~al.(2017)}]{van2017neural}
Van Den~Oord, A.; Vinyals, O.; et~al. 2017.
\newblock Neural discrete representation learning.
\newblock \emph{Advances in neural information processing systems}, 30.

\bibitem[{Van~Le and Kurosawa(2003)}]{van2003efficient}
Van~Le, T.; and Kurosawa, K. 2003.
\newblock Efficient public key steganography secure against adaptively chosen
  stegotext attacks.
\newblock \emph{Cryptology ePrint Archive}.

\bibitem[{Vaswani et~al.(2017)Vaswani, Shazeer, Parmar, Uszkoreit, Jones,
  Gomez, Kaiser, and Polosukhin}]{vaswani2017attention}
Vaswani, A.; Shazeer, N.; Parmar, N.; Uszkoreit, J.; Jones, L.; Gomez, A.~N.;
  Kaiser, {\L}.; and Polosukhin, I. 2017.
\newblock Attention is all you need.
\newblock \emph{Advances in neural information processing systems}, 30.

\bibitem[{Wang et~al.(2020)Wang, Li, Zhang, Yu, Liu, and Yu}]{wang2020bbc}
Wang, Y.; Li, W.; Zhang, W.; Yu, X.; Liu, K.; and Yu, N. 2020.
\newblock BBC++: Enhanced block boundary continuity on defining non-additive
  distortion for JPEG steganography.
\newblock \emph{IEEE Transactions on Circuits and Systems for Video
  Technology}, 31(5): 2082--2088.

\bibitem[{Wang et~al.(2019)Wang, Zhang, Li, Yu, and Yu}]{wang2019non}
Wang, Y.; Zhang, W.; Li, W.; Yu, X.; and Yu, N. 2019.
\newblock Non-additive cost functions for color image steganography based on
  inter-channel correlations and differences.
\newblock \emph{IEEE Transactions on Information Forensics and Security}, 15:
  2081--2095.

\bibitem[{Weng et~al.(2022)Weng, Chen, Yu, and Sun}]{weng2022lightweight}
Weng, S.; Chen, M.; Yu, L.; and Sun, S. 2022.
\newblock Lightweight and effective deep image steganalysis network.
\newblock \emph{IEEE Signal Processing Letters}, 29: 1888--1892.

\bibitem[{Xia et~al.(2022)Xia, Zhang, Yang, Xue, Zhou, and Yang}]{xia2022gan}
Xia, W.; Zhang, Y.; Yang, Y.; Xue, J.-H.; Zhou, B.; and Yang, M.-H. 2022.
\newblock Gan inversion: A survey.
\newblock \emph{IEEE transactions on pattern analysis and machine
  intelligence}, 45(3): 3121--3138.

\bibitem[{Yang et~al.(2018)Yang, Chen, Zhang, and Yu}]{yang2018provably}
Yang, K.; Chen, K.; Zhang, W.; and Yu, N. 2018.
\newblock Provably secure generative steganography based on autoregressive
  model.
\newblock In \emph{International Workshop on Digital Watermarking}, 55--68.
  Springer.

\bibitem[{Yang et~al.(2023)Yang, Chen, Zeng, Zhang, and Yu}]{yang2023provably}
Yang, Z.; Chen, K.; Zeng, K.; Zhang, W.; and Yu, N. 2023.
\newblock Provably secure robust image steganography.
\newblock \emph{IEEE Transactions on Multimedia}.

\bibitem[{Zhang et~al.(2022)Zhang, Gu, Zhang, Bao, Chen, Wen, Wang, and
  Guo}]{zhang2022styleswin}
Zhang, B.; Gu, S.; Zhang, B.; Bao, J.; Chen, D.; Wen, F.; Wang, Y.; and Guo, B.
  2022.
\newblock Styleswin: Transformer-based gan for high-resolution image
  generation.
\newblock In \emph{Proceedings of the IEEE/CVF conference on computer vision
  and pattern recognition}, 11304--11314.

\bibitem[{Zhang et~al.(2021)Zhang, Yang, Yang, and Huang}]{zhang2021provably}
Zhang, S.; Yang, Z.; Yang, J.; and Huang, Y. 2021.
\newblock Provably Secure Generative Linguistic Steganography.
\newblock In \emph{Findings of the Association for Computational Linguistics:
  ACL-IJCNLP 2021}, 3046--3055.

\end{thebibliography}

\end{document}